\documentclass[%
    reprint,
superscriptaddress,
 nofootinbib,
 amsmath,amssymb,
 aps,
 pra,
 onecolumn,
 notitlepage
]{revtex4-1} 

\usepackage[centering,hmargin=2cm,vmargin=2.5cm]{geometry}

\usepackage{graphicx}	
\graphicspath{{figures/}{./}} 
\usepackage{dcolumn}	
\usepackage{bm}			
\usepackage{hyperref}
\usepackage{ulem}
\usepackage{caption}
\usepackage{subcaption}

\usepackage{braket}
\newcommand{\ketbra}[2]{\vert #1 \rangle\langle #2 \vert}
\usepackage{listings}
\usepackage{color}
\usepackage{realboxes}
\usepackage{enumitem}   
\setlist{nolistsep}

\usepackage{float}
\makeatletter
\let\newfloat\newfloat@ltx
\makeatother
\usepackage{algorithm}
\usepackage{algorithmicx}
\usepackage{algpseudocode}
\algnewcommand\algorithmicswitch{\textbf{switch}}
\algnewcommand\algorithmiccase{\textbf{case}}
\algnewcommand\algorithmicassert{\texttt{assert}}
\algnewcommand\Assert[1]{\State \algorithmicassert(#1)}%

\definecolor{codegreen}{rgb}{0,0.6,0}
\definecolor{codegray}{rgb}{0.5,0.5,0.5}
\definecolor{codepurple}{rgb}{0.58,0,0.82}
\definecolor{backcolor}{rgb}{0.9,0.9,0.88}
\definecolor{mygray}{rgb}{0.8,0.8,0.8}

\begin{document}

\title{Solving Quadratic Unconstrained Binary Optimization with divide-and-conquer and quantum algorithms}

\author{Gian Giacomo Guerreschi}
 \email{gian.giacomo.guerreschi@intel.com}
 \affiliation{Intel Labs, Santa Clara, CA 95054, USA}


\date{\today}


\begin{abstract}
Quadratic Unconstrained Binary Optimization (QUBO) is a broad class of optimization problems with many practical applications. To solve its hard instances in an exact way, known classical algorithms require exponential time and several approximate methods have been devised to reduce such cost. With the growing maturity of quantum computing, quantum algorithms have been proposed to speed up the solution by using either quantum annealers or universal quantum computers. Here we apply the divide-and-conquer approach to reduce the original problem to a collection of smaller problems whose solutions can be assembled to form a single Polynomial Binary Unconstrained Optimization instance with fewer variables. This technique can be applied to any QUBO instance and leads to either an all-classical or a hybrid quantum-classical approach. When quantum heuristics like the Quantum Approximate Optimization Algorithm (QAOA) are used, our proposal leads to a double advantage: a substantial reduction of quantum resources, specifically an average of $\sim 42\%$ fewer qubits to solve MaxCut on random 3-regular graphs, together with an improvement in the quality of the approximate solutions reached.
\end{abstract}


\maketitle




\section{Introduction}
\label{sec:introduction}

Many problems of practical interest are phrased in terms of the optimization of binary-variable models with a quadratic cost function, a class of problems usually referred to as Quadratic Unconstrained Binary Optimization (QUBO). Its application spans tasks as diverse as resources allocation, clustering, set partitioning, facility locations, various forms of assignment problems, sequencing and ordering problems \cite{Kochenberger2014, Glover2018a}. In physics, QUBO corresponds to the Ising model describing spins with 2-body interactions \cite{Baxter1982}. In addition, a few NP-hard problems are naturally expressed as QUBO and central to the theory of computational complexity \cite{Feige1995, Lucas2014}. For its relevance, a vast number of algorithms have been developed to solve QUBO instances, ranging from exact solvers, to approximate solvers, to heuristics without performance guarantee \cite{Dunning2018}. Recently, quantum algorithms have joined the competition.

Quantum computing is a new paradigm to manipulate information based on quantum mechanics. It allows updating an exponentially large amount of information with a single operation and in ways that provide speedup for specific applications. Noticeable algorithms focus on integer factorization \cite{Shor1999}, database search \cite{Grover1997}, and simulation of physical \cite{Abrams1999} and chemical systems \cite{Cao2019}. The application to QUBO problems became central to quantum computing when adiabatic quantum optimization was proposed around the turn of the century \cite{Kadowaki1998, Farhi2001a} and quantum annealers became the first large-scale quantum devices a dozen years later \cite{Boixo2014, Santra2014, Venturelli2015}. With the accelerating development of universal quantum computers, the Quantum Approximate Optimization Algorithm (QAOA) has been proposed to solve binary optimization by encoding the approximate solution as a variationally-optimized quantum state \cite{Farhi14_qaoa_orig, Wecker16_training, Guerreschi17_qaoa_opt, Zhou2018, Shaydulin2019a, Streif2019a}.

QAOA is a leading candidate to achieve quantum advantage by solving a practical problem faster or with larger accuracy than classical alternatives. This expectation is based on the fact that QAOA requires relatively shallow circuits suitable for non-error corrected devices, and its variational nature mitigates the effect of systematic errors. Increasingly larger and more sophisticated experiments have been performed in the last few years \cite{Otterbach2017a, Mueller2018, Pichler2018a, Pagano2020, google2020b}. However, the number of qubits composing near-term quantum computers is expected to be one of the most severe limitations for running algorithms, arguably the most severe together with coherence time. Earlier estimates suggest that several hundreds of qubits are required for QAOA to compete with classical solvers \cite{Guerreschi2019} or to generate outcome distributions beyond classical simulability \cite{Dalzell2020}. To the best of our knowledge, no approach has been considered so far to reduce the number of qubits required by QAOA. An interesting approach named Recursive QAOA \cite{Bravyi2020} reduces the number of qubits during the optimization of the circuit by identifying the strongest-correlated spins, but with this technique it is the size of the initial circuit that is the limiting factor in terms of number of qubits.

In this work we explore a divide-and-conquer approach with the goal of reducing the number of variables of QUBO instances. Since the number of variables is typically a key factor in determining the cost of solving a binary optimization instance, there is the possibility that the reduced instance may also be easier to solve than the original one. While this additional benefit would be expected when comparing instances drawn from the same complexity class, we observe that the modified instances belong to a broader class of optimization problems called Polynomial Unconstrained Binary Optimization (PUBO). As the name suggests, this class includes instances whose cost function has polynomial degree beyond quadratic and it is not \textit{a priori} clear whether they are harder to solve than QUBO instances of the same size.

The technique we propose is based on community detection algorithms together with a novel improvement specialized for variable reduction. Our proposal can be used as the initial step of an all-classical approach, or as part of a mixed classical-quantum one. The latter takes advantage of the natural flexibility of QAOA which can be readily adapted from the solution of QUBO to that of PUBO. We observe an interesting situation if we compare the benefits provided by the divide-and-conquer step either to an exact classical solver or to QAOA. Our numerical experiments suggest that only the quantum heuristic algorithm takes advantage of the variable elimination.
Specifically, we consider the graph partition problem called Max-Cut and two sets of instances corresponding to random 3- and 4-regular graphs. We show that the reduced instances require $\sim 42\%$, respectively $\sim 22\%$ fewer variables. However the more compact formulation does not correspond to a faster solution when the exact solver \texttt{akmaxsat} \cite{Kugel2012} is used. On the contrary, it translates to better approximate solutions when QAOA is considered.

The paper is organized as follows. Section~\ref{sec:background} covers the technical background: we introduce QUBO and its generalization to PUBO, then we explicitly rephrase Max-Cut as QUBO to set the stage for later results.
We also discuss QAOA and how it can be used to approximately solve PUBO problems. All of these are known results.
In Section~\ref{sec:div-n-con}, we explain our proposal of using community detection algorithms to reduce the original problem to one with fewer variables.
In Section~\ref{sec:results}, we quantify the effect of the divide-and-conquer approach in terms of variable elimination for the reduced instance. We also compute the cost of an all-classical implementation and compare it with the situation in which QAOA is applied to the reduced instance. This Section demonstrates a double advantage of using divide-and-conquer together with QAOA: fewer qubits are required and higher approximation ratio is reached.
At the end we draw conclusions and present some open questions.

\section{Background and definitions}
\label{sec:background}

\subsection{QUBO and its generalization to PUBO}

As the name suggest, QUBO represents optimization problems in which a quadratic function on $N$ binary variables has to be minimized over all possible $2^N$ assignments of its variables. We refer to the function to minimize as the cost function or energy, and it can be written as:
\begin{equation}
\label{eq:qubo}
	E_\text{qubo}(\vec s) = \alpha \,+\, \sum_{i} \alpha_i \; s_i \,+\, \sum_{\langle i_1, i_2 \rangle} \alpha_{i_1, i_2} \; s_{i_1} s_{i_2} \; .
\end{equation}
where $\vec s=(s_1, s_2, \dots, s_N)$ represents the assignment of $N$ spins and $\langle i_1, i_2 \rangle$ indicates a pair of spins, those with indices $i_1, i_2$. Spins have values $s_i \in \{+1, -1\}$, and coefficients $\alpha, \,\alpha_i, \,\alpha_{i_1, i_2}$ are real.
Here and in the following we will use the term variable and spin interchangeably.

It is useful to introduce a generalization of QUBO to Polynomial Unconstrained Binary Optimization (PUBO). This larger class relaxes the constraint of having terms involving at most two spins. Any PUBO cost function can be written as:
\begin{equation}
\label{eq:pubo}
	E_\text{pubo}(\vec s) = \sum_{k=0}^N \, \sum_{\langle i_1, i_2, \dots, i_k \rangle} \alpha_{i_1, i_2, \dots, i_k} \; s_{i_1} s_{i_2} \dots s_{i_k} \; ,
\end{equation}
where $\langle i_1, i_2, \dots, i_k \rangle$ indicates a group of $k$ spins (those with indices $i_1, i_2, \dots, i_k$) and coefficients $\alpha_{i_1, i_2, \dots, i_k}$ are real. While the general formulation has $2^N$ terms, PUBOs of practical interest usually have terms involving only $k\leq \tilde k$ spins for $\tilde k=\mathcal{O}(1)$, and therefore only a polynomial number of coefficients are non-zero.

\subsection{Max-Cut: a graph partition problem expressed as QUBO}

Given a graph, consider the problem of assigning one of two colors to each of the vertices. This creates a partition of the graph into two sets of vertices. Every edge that connects vertices of different color is considered ``cut'' by the partition. The Max-Cut problem asks to find the maximum number of edges that can be cut by any of the possible partitions. Despite its apparent simplicity, Max-Cut is a NP-complete problem and no known algorithm, either classical or quantum, can solve its worst case in polynomial time \cite{Karp1972, Dunning2018}.

Max-Cut instances are naturally described in terms of QUBO. Assign to vertex $i$ the spin $s_i\in\{+1,-1\}$ and consider the partition defined by the set of spins up, those with $s_i=+1$, and spins down, with $s_i=-1$. The quantity $\tfrac{1 - s_i s_j}{2}$ is null when $s_i=s_j$ and equal to 1 when $s_i=-s_j$, effectively determining if an edge between vertex $i$ and $j$ is cut. The cost function can be expressed as ``minus'' the number of cut edges:
\begin{equation}
\label{eq:original-maxcut-energy}
	E_\text{maxcut}(\vec s) = - \sum_{\langle i, j \rangle} \frac{1 - s_i s_j}{2}
					   = \sum_{\langle i, j \rangle} \frac{s_i s_j}{2} - \frac{M}{2} \; ,
\end{equation}
with the summation being on the edges of the graphy to partition and $M$ being the total number of edges.
It is clear by comparison with Eq.~\eqref{eq:qubo} that $E_\text{maxcut}(\vec s)$ is of QUBO form.

By associating a weight to every edge, any QUBO without linear terms can be formulated as weighted Max-Cut in which the goal is maximizing the weight of cut edges instead of their number.
For ease of description and visualization, we will present our divide-and-conquer approach in the context of Max-Cut by using terms and concepts from graph theory. The extension to QUBO with linear terms is straightforward and does not present any computational subtlety.

\subsection{Quantum Approximate Optimization Algorithm}
\label{sec:qaoa}

The Quantum Approximate Optimization Algorithm (QAOA) is a hybrid quantum-classical algorithm designed to find approximate solutions of combinatorial optimzation problems by variationally improving quantum circuits \cite{Farhi14_qaoa_orig, Wecker16_training, Guerreschi17_qaoa_opt, Zhou2018, Shaydulin2019a, Streif2019a}. QAOA is regarded as a strong candidate for near-term applications since it uses relatively shallow quantum circuits, its variational nature provides robustness to systematic errors, and the classical alternatives are costly even for the solution of relatively small instances of combinatorial problems. QAOA can be applied to any binary optimization and thus, in the language of this work, to any PUBO.

Central to QAOA are two quantum Hamiltonians, or quantum energy functions. The first is a direct translation of the classical function to minimize, $E_\text{pubo}(\vec s)$, obtained by substituting the spin $s_i$ with the quantum operator $\hat Z_i$ (i.e. the Pauli $Z$ matrix on qubit $i$). We denote this cost Hamiltonian by $\hat H_\text{pubo}$. The second is a driver Hamiltonian, required to be non-commuting with $\hat H_\text{pubo}$ and typically chosen to correspond to the homogeneous field in the $X$ direction. Formally:
\begin{align}
	\hat H_\text{pubo}   &= \sum_{k=0}^N \, \sum_{\langle i_1, i_2, \dots, i_k \rangle} \alpha_{i_1, i_2, \dots, i_k} \; Z_{i_1} \hat Z_{i_2} \dots \hat Z_{i_k} \; ,\\
	\hat H_\text{driver} &= \sum_{i=1}^N \, \hat X_i \; ,
\end{align}
where $\hat Z_i$ and $\hat X_i$ are the Pauli matrices $Z$ and $X$ of qubit $i$. The form of $\hat H_\text{pubo}$ is completely analogue to Eq.~\eqref{eq:pubo}.

The above Hamiltonians are used to characterize the quantum circuit of QAOA. Specifically, the circuit is formed by alternating $p$ applications of $\hat U(\gamma_k)=\exp(-i \gamma_k \hat H_\text{pubo})$ and $\hat V(\beta_k)=\exp(-i \beta_k \hat H_\text{driver})$. Parameters $\gamma_k$ and $\beta_k$ may differ for each application $k=1,2,\dots,p$. The QAOA quantum circuit can be seen as a preparation of the parametric state $\ket{\psi(\vec \gamma, \vec \beta)}$:
\begin{equation}
	\ket{\psi(\vec \gamma, \vec \beta)} = \underbrace{\hat V(\beta_p) \hat U(\gamma_p)}_\text{last layer}
		\dots \hat V(\beta_2) \hat U(\gamma_2) \,
		\underbrace{\hat V(\beta_1) \hat U(\gamma_1)}_\text{first layer} \ket{\psi(0)}
\end{equation}
where $\vec \gamma = (\gamma_1, \dots, \gamma_p)$ and $\vec \beta = (\beta_1, \dots, \beta_p)$. When measured in the $Z$ basis $\{\ket{\vec s}\}_{\vec s \in \{1,-1\}^N}$, the state returns an assignment $\vec s$ of the original $N$ spins of the PUBO instance. The assignment is not unique, but determined by the probability distribution $P(\vec s) = \left| \braket{\vec s \vert \psi(\vec \gamma, \vec \beta)} \right|^2$. By varying parameters $(\vec \gamma, \vec \beta)$ using a classical optimizer, the distribution can be changed to increase the probability of measuring assignments corresponding to low values of $E_\text{pubo}$. The figure of merit of the parameters' optimization is usually chosen to be the expectation value of the energy over the distribution $P(\vec s)$, namely $\braket{\psi(\vec \gamma, \vec \beta) \vert \hat H_\text{pubo} \vert \psi(\vec \gamma, \vec \beta)}$. Other choices are possible \cite{Barkoutsos2020, Larkin2020a}. When the exact solution is known, results are typically reported in term of the approximation ratio:
\begin{equation}
\label{eq:app-ratio}
	f_p(\vec \gamma, \vec \beta) = \frac{\braket{\psi(\vec \gamma, \vec \beta) \vert \hat H_\text{pubo} \vert \psi(\vec \gamma, \vec \beta)}}
	                                    {\min_{\vec s} \; E_\text{pubo} (\vec s)}
\end{equation}


One of the major limitations of quantum hardware at this stage of development is the number of qubits composing the system. In the next Section, we introduce techniques to reduce the need of qubits for QAOA well below the number $N$ of original variables of Max-Cut or QUBO.

\section{Divide-and-Conquer applied to QUBO}
\label{sec:div-n-con}

\subsection{Community detection}

Given a QUBO instance, consider its quadratic terms only and neglect for the moment the constant and linear terms. The interaction pattern among variables can be visualized as a graph in which each vertex correspond to a spin and each edge to a quadratic term. The hardness to solve the QUBO instance is related to this interaction graph and typical benchmarks require solving non-planar graphs with random edges. A powerful method to analyze the structure of a graph is to divide its vertices into communities. A community is a subset of vertices that are strongly connected among themselves while exhibiting a relatively small number of across-community edges. The communities are disjoint and their union includes all vertices of the original graph. This suggests a way to divide the original problem into a set of sub-problems by considering the original QUBO instance restricted to each community. Finally, one needs to consider the original inter-community edges while patching together the partial solutions.

There is not a single way to define the most desirable division in communities, but standard approaches tend to group vertices in such a way that the number of inter-community edges is minimized, as quantified by the modularity \cite{Fortunato2016}. In practice, all metrics include terms that oppose the tendency of collecting all vertices into a single community, and most methods return a small number of communities. In our case, we want to maximze the benefit of using the quantum heuristic QAOA to solve the reduced PUBO and, specifically, we want to minimize the number of qubits needed to encode the original problem. As we will explain in the next paragraphs, this is obtained not when the number of inter-community edges is minimized, but rather when the number of vertices with inter-community edges is minimized. Despite being related, we will show how an ad-hoc modification of standard community detection algorithms leads to further, substantial, qubit reduction.

\begin{figure}[b]
  \centering
  \includegraphics[width=0.25\textwidth]{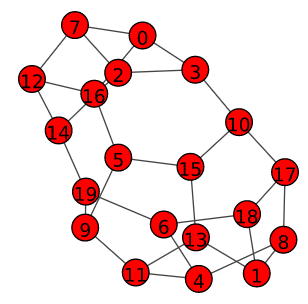}
  \hspace{5mm}
  \includegraphics[width=0.25\textwidth]{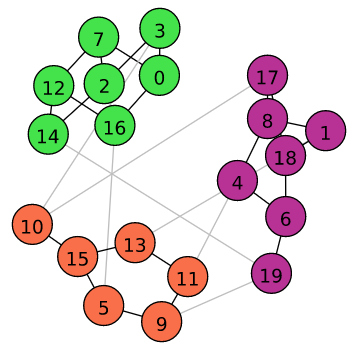}
  \hspace{5mm}
  \includegraphics[width=0.25\textwidth]{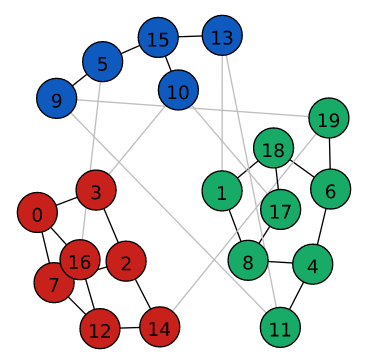}
  \caption{Comparison between two different grouping of a 3-regular graph with $N=20$ vertices into three communities. On the left, the original graph. In the center, the grouping is provided by a standard community detection algorithm used as baseline. Specifically, it is the algorithm \texttt{community\_multilevel} from iGraph \cite{igraph} which has the goal of maximizing the graph modularity (strongly related to minimizing the number of across-community edges). On the right, the grouping after our ad-hoc algorithm aiming at reducing the number of vertices with across-community edges, which we call boundary vertices. In fact, it is the latter number that determines how many qubits are needed when encoding the MaxCat problem of the original graph into a many-body Hamiltonian. The number of qubits for this graph instance are reduced from 20 to 12 to 11.}
  \label{fig:community-detection}
\end{figure}

The community detection algorithm that we adopt as baseline is \texttt{community\_multilevel} as implemented in the iGraph library \cite{igraph}. It is a bottom-up algorithm that starts from a fine-grained view of the graph and moves towards a coarse-grained view. Initially every vertex is its own community, then vertices are moved between communities to maximize the overall modularity score. When no single vertex movement can increase the modularity, every community is shrank to a single vertex and the process restarts. The algorithm stops when neither shrinking nor vertex movement further improve the modularity. The final result assigns a membership value $c_i \in C$ to every vertex $i$, indicating the community it belongs to. We denote the set of communities with $C$.

Our ad-hoc improvement starts with the communities identified by \texttt{community\_multilevel}. It uses a different score based on the concept of ``boundary'' vertices. Denoting the set of vertices by $S$ (a reminder that they play the role of spins in the QUBO formulation of Max-Cut), divide it in subsets according to the community membership. We define two subset per community, $B_c$ and $T_c$, with $c$ being the community index:
\begin{align}
B_c &= \{s_i \,|\, c_i = c \land (\exists \langle i, j \rangle \in \text{Edges} \text{ s.t. } c_i\neq c_j) \} \\
T_c &= \{s_i \,|\, c_i = c \} \setminus B_c \; .
\end{align}
These sets are pairwise disjoint and their union is $S$. Intuitively, $B_c$ is the set of boundary vertices of community $c$, i.e. those vertices that have at least one edge connecting them to vertices of different communities. $T_c$ is the complementary set restricted to community $c$, and represents the ``core'' vertices of the community. One can think of the boundary vertices from all communities as forming the set $B=\bigcup_{c\in C} B_c$.

Our ad-hoc improvement moves vertices between communities with the scope of minimizing the score $g$. No multi-level strategy is used, so the algorithm stops when updating any single membership $c_i$ does not decrease $g$.
The score we use is $g(C) = \max \{ |B|,  \max_c (|B_c|+|T_c|) \}$, the maximum between the size of the overall boundary and the size of every community. As we will see in the next Section, $|B|$ corresponds to the number of qubits required to solve the original QUBO instance following the divide-and-conquer approach. The second term counteracts the tendency to form very large communities and avoid the risk that the computational bottlenecks becomes solving the subproblem for the largest community. We further elaborates on the reasons of this choice in Section~\ref{sec:derivation_pubo}. To help visualization, Fig.~\ref{fig:community-detection} shows a 3-regular graph with 20 vertices and two ways of dividing it into three communities.

\subsection{Divide-and-conquer}

Using insight from community detection, we can write the energy function from Eq.~\eqref{eq:qubo} as: 
%
\begin{equation}
\label{eq:Equbo1}
E_\text{qubo}(\vec s)
	= \sum_{c\in C} \underbrace{ \left[ \sum_{i} \delta_{c_i,c}\,\alpha_i\,s_i \,+\, \sum_{\langle i, j \rangle} \alpha_{i,j} \,\delta_{c_i,c}\,\delta_{c_j,c} \;s_i\,s_j \right]}
	_{\text{contribution from community } c}
	\,+\, \underbrace{\left[ \alpha \,+\, \sum_{\langle i, j \rangle} \alpha_{i,j} \,\big( 1-\delta_{c_i,c_j} \big) \;s_i\,s_j \right]}_{\text{across-community contributions}} \; ,
\end{equation}
where $C$ represents the set of communities, $c_i$ the membership of vertex $i$, and $\delta_{\star, \star}$ the Kronecker symbol. The first term is the sum of intra-community interactions, while the second term represents the contribution from across-community interactions. By convention, we include all linear terms as part of the intra-community term and the constant term as part of the across-community term.

Recall that $B_c$ corresponds to the boundary spins of community $c$, \textit{i.e.} those that have connections with vertices in other communities, while $T_c$ are the remaining non-boundary spins of the community, also called its core. The energy $E_\text{qubo}(\vec s)$ can therefore be written as:
\begin{align}
\label{eq:Equbo2}
	E_\text{qubo}(\vec s) = \sum_{c\in C} E^{(c)}_\text{intra-comm}(\vec b_c, \vec t_c) \,+\, E_\text{across-comm}(\vec b) \; ,
\end{align}
where both $E^{(c)}_\text{intra-comm}$ and $E_\text{across-comm}$ have at most quadratic terms.
We use notation $\vec t_c, \,\vec b_c, \,\vec b$ to represent the assignment $\vec s$ restricted to spins in $T_c, \,B_c, \, B$ respectively.

The ultimate goal is finding or approximating the ground state energy of $E_\text{qubo}(\vec s)$. To this extent we notice that the lowest part of the $E_\text{maxcut}$ spectrum is reproduced by an energy function on boundary spins alone. This can be achieved by eliminating the explicit dependency from the core variables of community $c$. Consider substituting the contribution $E_\text{intra-comm}(\vec  b_c, \vec t_c)$ with:
\begin{align}
\label{eq:intra-comm-energy}
	\tilde{E}^{(c)}_\text{intra-comm}(\vec b_c) = \min_{\vec t_c \in \{+1, -1\}^{|T_c|}} E^{(c)}_\text{intra-comm}(\vec b_c, \vec t_c) \; .
\end{align}
The original QUBO energy is then written as a function of boundary variables alone:
\begin{align}
\label{eq:reduced-energy}
	\tilde{E}(\vec b) = \sum_{c \in C} \tilde{E}^{(c)}_\text{intra-comm}(\vec b_c) \,+\, E_\text{across-comm}(\vec b) \; .
\end{align}

Our proposal is to solve for the ground state energy of $\tilde{E}$ using classical solvers or quantum algorithms. A few considerations:
\begin{itemize}
	\item the ground state energy of $\tilde{E}$ exactly corresponds to that of $E_\text{qubo}$;
	\item the maximization procedure to compute $\tilde{E}^{(c)}_\text{intra-comm}(\vec b_c)$ typically takes a negligible effort w.r.t. the solution of the full problem since it requires solving similar, much smaller problems with $|T_c|$ spins instead of $N$ and the cost is exponential in the number of spins. However, one has to solve one of the smaller problems for each assignment of the spins in $B_c$;
	\item $\tilde{E}^{(c)}_\text{intra-comm}(\vec b_c)$ is an energy function with polynomial degree at most $|B_c|$;
	\item when considering Max-Cut, due to the $\mathbb{Z}_2$ symmetry of the original energy $E_\text{qubo}$, every term of the energy function $\tilde{E}^{(c)}_\text{intra-comm}$ involves only an even number of variables;
	\item we provide a constructive way to build $\tilde{E}^{(c)}_\text{intra-comm}$ in the Section~\ref{sec:derivation_pubo}, following the approach presented in \cite{Sawaya2020} with a more efficient implementation.
\end{itemize}

\section{Results}
\label{sec:results}

\subsection{Community detection and variable elimination}
\label{sec:variable-elimination}

\begin{figure}[b]
  \centering
  \includegraphics[width=.95\textwidth]{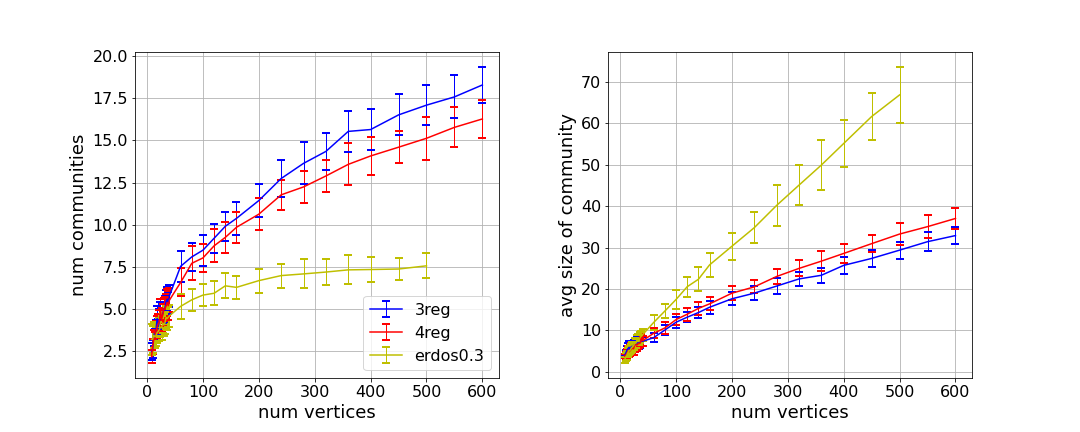}
  \caption{\textbf{(Left)} Number of communities as a function of the number of vertices of the original graph. \textbf{(Right)} Average number of vertices in a community. In both panels, each datapoint corresponds to the average of 100 instances and vertical bars represent one standard deviation. Three classes of graphs are shown: in blue random 3-regular graphs, in red random 4-regular graphs, and in yellow Erdos-Renyi graphs with $p_E=0.3$.}
  \label{fig:community-size}
\end{figure}

We begin by quantifying the outcome of the community detection step in terms of the number and size of the subproblems it creates. We consider interaction graphs belonging to two important classes of random graphs: $k$-regular graphs in which each vertex has exactly $k$ edges, and Erdos-Renyi graphs in which an edge between any pair of vertices is present with probability $p_E$. For this study we consider regular graphs with $k=3,4$ and Erdos-Renyi with $p_E=0.3$. In particular, Max-Cut on random 3-regular graphs has been widely studied both classically \cite{Goemans1995,Halperin2004} and as benchmark of QAOA \cite{Farhi14_qaoa_orig, Zhou2018, Guerreschi2019}.

Fig.~\ref{fig:community-size} shows the slow increase in the number of communities identified by the community detection algorithm \texttt{community\_multilevel} from iGraph \cite{igraph}. The average size of the communities is simply the number of communities divided by the number of vertices of the original graph. We observe a markedly different behavior between $k$ regular graphs and Erdos-Renyi graphs that we associate with the number of edges: while $k$-regular graphs of $N$ vertices have exactly $kN/2$ edges, Erdos-Renyi graphs have an expected number of edges equal to $p_E N(N-1)/2$. Erdos-Renyi graphs are therefore much more dense than $k$-regular ones.
For each graph class, we report a single dataset corresponding to the result of a standard community detection algorithm improved by our ad-hoc post-process. In terms of number of communities, the post-process does not change the number of communities apart from small-size effects for graphs with few tens of vertices. However, the post-process leads to measurable changes in the number of boundary vertices, as we analyze next.

\begin{figure}[t]
  \centering
  \includegraphics[width=.95\textwidth]{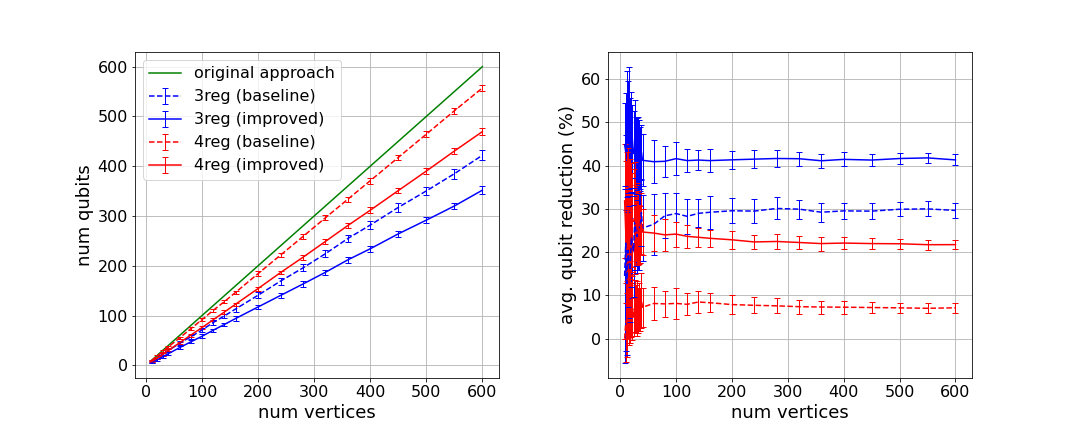}
  \caption{\textbf{(Left)} Number of qubits needed to run QAOA for Max-Cut instances following the standard QAOA approach and for our proposal. Two classes of graphs are shown: in blue random 3-regular graphs, and in red random 4-regular graphs. The baseline is obtained by performing community detection with a modularity-based algorithm, the \texttt{community\_multilevel} from iGraph \cite{igraph} (dash lines). The improved results are obtained using our specialized post-process (full lines). \textbf{(Right)} Qubit reduction quantified as percentage of $(1-|B|/N)$. It demonstrates a qubit reduction of more than 40\% with respect to the standard approach for random 3-regular graphs. In both panels, each datapoint corresponds to the average of 100 instances and vertical bars represent one standard deviation. Results for Erdos-Renyi graphs with $p_E=0.3$ show a limited impact of the divide and conquer approach and no qubit reduction is present when the original graph has 40 or more vertices.}
  \label{fig:qubit-reduction}
\end{figure}

We compute the total number of boundary variables, i.e. $|B|$, and relate it to the number of qubits required to apply QAOA to the original problem following the divide-and-conquer approach we are proposing. The number of available qubits is expected to be one of the major limitations of near-term devices. Reducing the qubit requirements of an algorithm would provide significant benefits both in terms of when the algorithm can be realized in practice (devices with fewer qubits are expected to be developed before devices with more qubits) and of the effective decoherence rate (smaller for a register with fewer qubits of given quality). Remarkably, in Section~\ref{sec:divncon-qaoa} we also show that QAOA returns solutions with higher approximation ratio when adopting the divide and conquer approach.

The reduction in the number of required qubits is quantified in Fig.~\ref{fig:qubit-reduction}, in which the original approach of one qubit per graph vertex (equivalently per binary variable) is compared with the divide-and-conquer approach based on standard community detection algorithms and our ad-hoc improvement. For random 3-regular graphs, the required number of qubits is reduced by $\sim 30\%$ and $\sim 42\%$ over a large range of graph sizes. For random 4-regular graphs the reduction is $\sim 8\%$ and $\sim 22\%$ respectively.

\subsection{Fully-classical exact solver}
\label{sec:akmaxsat-only}

The divide-and-conquer approach reduces the size of the QUBO instance to be solved at the cost of solving a large number of partially quenched instances. These subinstances correspond to the restriction of the original QUBO to single communities of vertices whose boundary is quenched to specific assignments. An important question is whether this approach provides any advantage in terms of a faster solution of the original instance. To address this point, we identify the total cost of the divide-and-conquer strategy as the sum of four terms:
\begin{enumerate}
\item[(1)] Divide original graph into communities using standard algorithms for community detection. Optionally, update the division in communities by minimizing the number of vertices with inter-community connections instead of the number of edges between communities.
\item[(2)] Solve the \textit{partially quenched} QUBO subinstances for every community and every assignments of the boundary variables.
\item[(3)] Combine the partial solutions to create the reduced PUBO instance whose solution exactly corresponds to the solution of the original instance.
\item[(4)] Solve the reduced PUBO instance.
\end{enumerate}
In this work, we adopt \texttt{akmaxsat} \cite{Kugel2012} as the exact solver to be used in steps (2) and (4). \texttt{akmaxsat} is a state-of-the-art solver of the Max-SAT problem to which both QUBO and PUBO can be reduced. It has been previously used to benchmark both quantum annealers \cite{Santra2014} and QAOA algorithms \cite{Guerreschi2019, Larkin2020a}.

\begin{figure}[b]
  \centering
  \includegraphics[width=.55\textwidth]{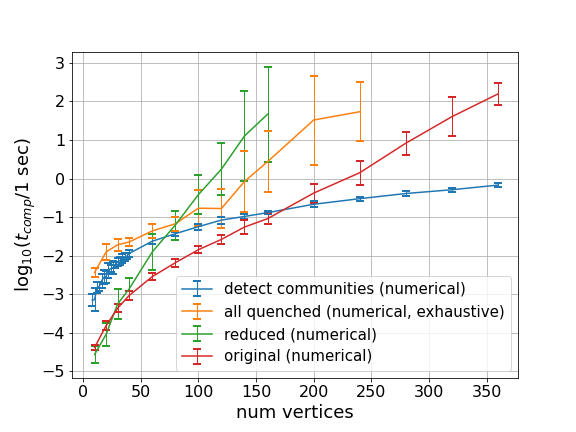}
  \caption{Comparison between the time taken to solve the original QUBO problem directly with \texttt{akmaxsat} (red) and the time to perform the steps of the divide-and-conquer approach. In blue, the time needed to detect the communities using \texttt{community\_multilevel} from iGraph and our \textit{ad-hoc} improvement. In orange, the time taken to solve all partially-quenched subinstances (exhaustively, one per boundary assignment of each community) together with the time needed to construct the reduced PUBO instance. In green, the time taken by \texttt{akmaxsat} to solve the reduced PUBO.
According to the list of the 4 terms characterizing the divide-and-conquer approach discuss in the main text, the orange line corresponds to the sum of terms (2) and (3), while the green line to term (4).
The x-axis corresponds to the number of vertices of the graph to partition, here corresponding to the number of spins of the original PUBO instance. The y-axis reports the $\log_{10}(\star)$ of the computational time in seconds. The datapoints are averages over 20 instances of Max-Cut for random 3-regular graphs, and the vertical bars correspond to one standard deviation.}
  \label{fig:akmaxsat}
\end{figure}

A few considerations on the steps listed above.
Related to (1), the cost of running the community detection algorithm scales favorably when compared to the rest of the protocol, even including the ad-hoc improvement. We confirm it with quantitative estimates below, but focus on the other contributions.
Concerning (2), the procedure detailed in Section~\ref{sec:derivation_pubo} requires the solution of an exponential number of subinstances: for each community $c$, we have $2^{|B_c|}$ way of constrained the boundary variables, and for each choice we have to solve a QUBO instance on $|T_c|$ core variables. This may be unnecessary when we are interested in finding good approximations of the solution to the original QUBO and not its global solution. Other, computationally less demanding, approaches are possible as we will discuss below. As the starting point, here we solve each subinstances using \texttt{akmaxsat}, an exact solver for SAT optimization.

About (3), the creation of the reduced PUBO instance also requires exponential time. The approach we consider to construct the contribution of community $c$ scales as $\mathcal{O}(|B_c|2^{|B_c|})$. In our study, we also need to express the PUBO in terms of the Conjunctive Normal Form (CNF) used by SAT optimizers. The method we follow to reduce PUBO to SAT instances is presented in Section~\ref{app:sec:pubo-to-sat} and requires $2^{k-1}$ clauses of $k$ variables each to represent a PUBO $k$-body term. When an approximate solution of (2) is sufficient, both the derivation cost and the number of clauses in the CNF can be significantly reduced.
Concerning (4), this is a natural place where classical solvers can be substituted by hybrid quantum-classical algorithms like QAOA. We explore this scenario in the next Section.

First of all, we consider whether an exhaustive implementation of step (2) leaves any possibility for the divide-and-conquer approach to be faster than the straightforward solution of the original instance. Fig.~\ref{fig:akmaxsat} reports (the base-10 logarithm of) the time to perform a few computations: in orange, the combined cost of step (2) and (3); in green the cost of step (4); in red, the cost of solving the original instance directly. While the red and green lines rely on a state-of-the-art SAT solver, the orange line is based on our own implementation and may thus be regarded as a upper bound. The difference between the red and green line is somewhat surprising since, for the same number of spins of the original instance (given by the horizontal axis), the reduced instance has actually fewer variables. However, the reduced instance has more clauses per variable and we observed in our numerical experiments that \texttt{akmaxsat} performs better for small density of clauses per variable. A different solver may alleviate the problem. 
%

\begin{figure}[b]
  \centering
  \includegraphics[width=.9\textwidth]{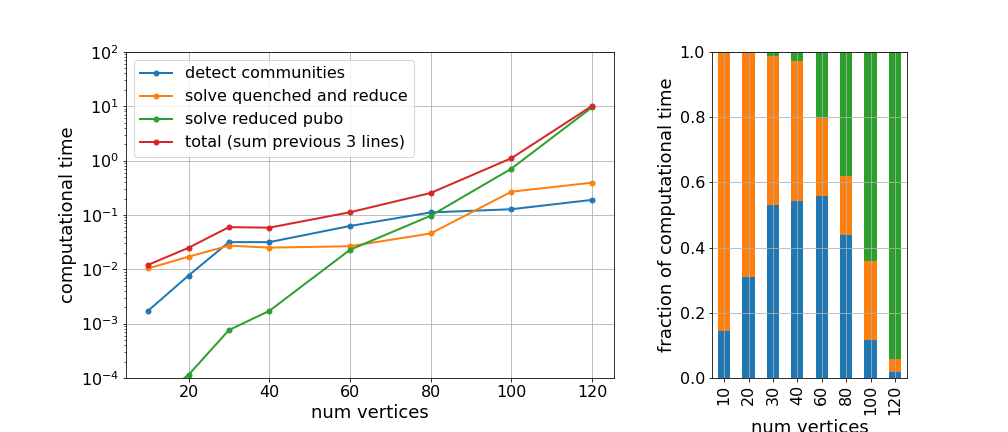}
  \caption{\textbf{(Left)} Computational time spent on the various step of the divide-and-conquer protocol described in the main text. According to the division in 4 steps, we have (1) in blue, (2)+(3) in orange, and (4) in green. The total is reported in red and substantially corresponds at $N = 120$ to the time required to solve the reduced instance. Each datapoint corresponds to the average over 20 instances of Max-Cut for random 3-regular graphs. Error bars are avoided for readability, but can be found in Fig.~\ref{fig:akmaxsat}. \textbf{(Right)} Stacked plot showing the time spent in the different steps of the protocol as fraction of the total.}
  \label{fig:dnc-cost}
\end{figure}

We then estimate the cost of the complete divide-and-conquer protocol in Fig.~\ref{fig:dnc-cost}. For small instances, the total cost is dominated by the solution of all subinstances. For large instances, starting at $N = 120$, the cost is dominated by the solution of the reduced instance. When compared with the cost of solving the original instance directly (see Fig.~\ref{fig:akmaxsat}), it is clear that the exhaustive application of the divide-and-conquer approach is not providing a computational advantage.

To increase the competitiveness of the divide-and-conquer approach it is fundamental to address the cost of steps (2-4).
One approach is solving the unconstrained subproblem once for each community, and fix the core variables to the corresponding best assignment. This reduces the cost of both (2) and (3): one needs to solve a single QUBO per community (despite on both boundary and core variables, instead of on core variables alone) and the reduced instance is obtained by fixing the assignment for the core variables. An additional advantage of this approximation is that the degree of the PUBO is actually not increased by the reduction and its derivation is straightforward. Therefore the reduced instance is also a QUBO and the divide-and-conquer procedure can be applied again. The main drawback is that the procedure does not guarantee that the global solution of the original instance is faithfully reproduced by the reduced instance. To mitigate this effect, once the assignment of the boundary variables is determined by the solution of the reduced instance, the core variables can be determined by solving the constrained subproblems and not assuming the assignment found in (2).

Concerning the steep rise of the cost of (4), we believe that this is due to the way we translate a PUBO instance to a SAT instance. In fact, the approach described in Section~\ref{app:sec:pubo-to-sat} requires $2^{k-1}$ clauses per $k$-body term. In addition, numerical experiments on different graph types (specifically fully connected and Erd\H{o}s-Renyi random graphs) suggest that \texttt{akmaxsat} performance degrades with increasing density of clauses per variable. It would be important to re-address the computational cost of step (4) using a solver specialized for PUBO instances.

\subsection{Quantum heuristic enhanced by classical divide-and-conquer}
\label{sec:divncon-qaoa}

One of the main advantages of the divide-and-conquer approach is that at every step of the protocol one has to solve PUBO instances with fewer spins than the original one. This is particularly suitable to quantum heuristics, like QAOA, whose application is currently limited by the number of qubits available to NISQ devices. In addition, smaller instances often translates in shorter quantum circuits and this may benefit the overall fidelity of the quantum computation.
The qubit reduction can be evaluated from the results of Section \ref{sec:variable-elimination}, and it reaches $\sim 40\%$ for random 3-regular graphs. 

Without including the effect of noise in our study, we are interested in whether our proposal improve the quality of the QAOA solution. We consider three cases: solving the original QUBO instance with QAOA, applying divide-and-conquer by using an exhaustive approach to solve all partially-quenched subinstances (as described in the previous Section) and then use QAOA for the reduced PUBO instance, or fixing the core spins of each community to their value for the best assignment of the subinstance (see discussion at the end of Section~\ref{sec:akmaxsat-only}) and then solve the reduced PUBO instance with QAOA.

\begin{figure}[b]
  \centering
  \includegraphics[width=.47\textwidth]{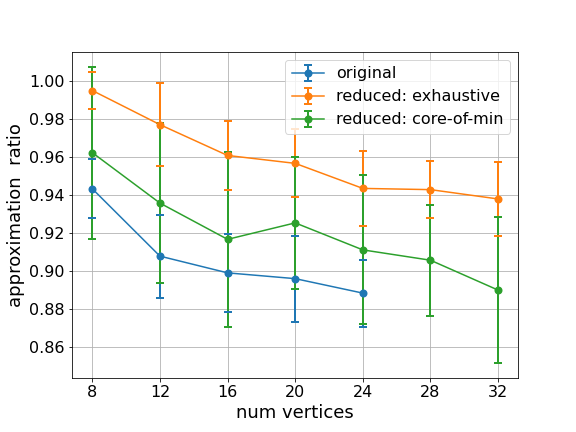}
  \includegraphics[width=.47\textwidth]{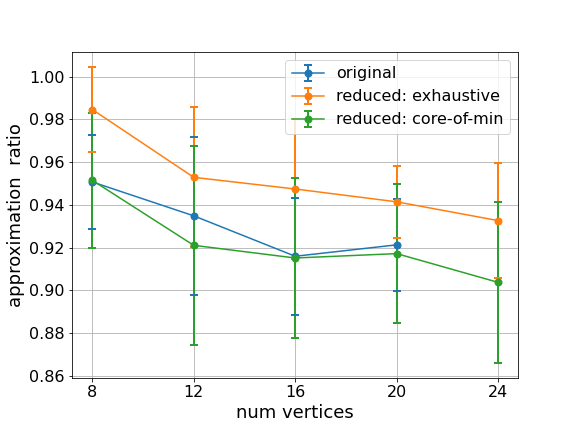}
  \caption{Performance of QAOA as quantified by the approximation ratio at the end of a global parameter optimization. Three situations are considered: solving the original QUBO (blue line), the reduced PUBO after exhaustive solution of all partially-quenched subinstances (orange), and the reduced PUBO obtained by fixing the core spins of each community separately (green). Each datapoint corresponds to the average over 20 instances of Max-Cut for random 3-regular \textbf{(left)} and 4-regular \textbf{(right)} graphs. The vertical bars correspond to one standard deviation.}
  \label{fig:qaoa-performance}
\end{figure}

The results are reported in Fig.~\ref{fig:qaoa-performance} as the blue, orange and green lines respectively. They suggest that optimizing the reduced PUBO problem via QAOA leads to better quality of the approximate solution compared to the optimization of the original QUBO. The quality of solution is provided in terms of the approximation ratio in Eq.~\eqref{eq:app-ratio}. For these experiments, we considered 20 instances of Max-Cut on random 3- and 4-regular graphs. Each instance (either the original or reduce one) was solved by optimizing the $(\vec \gamma, \vec \beta)$ parameters with a global approach known as APOSMM (Asynchronously Parallel Optimization Solver for Finding Multiple Minima) \cite{Larson2018} that coordinates multiple runs of the local optimizer \texttt{COBYLA} from SciPy package \cite{2020SciPy-NMeth}, starting from random initial parameter values.


Irrespective of the situation, the global optimization has a total budget of 10,000 function evaluations per instance and each QAOA circuit has depth $p=4$. As noted above, Fig.~\ref{fig:qaoa-performance} suggests that QAOA works more effectively for the reduced PUBO instances than for the original QUBO one. On the contrary, Fig.~\ref{fig:akmaxsat} shows that \texttt{akmaxsat} takes longer to solve the reduced instances instead of the original ones. It would be interesting to observe if this behavior is reproduced when other classical solvers are considered.

In addition, we observe the good performance of QAOA after non-exact reduction for random 3-regular graphs. By fixing the spins of each community core, we are giving up the exact correspondence between the original and reduced ground state energy, with the latter being an upper bound of the former. This is most probably the case in our study since $E^{(c)}_\text{intra-comm}$ has $\mathbb{Z}_2$ symmetry (see Eq.~\eqref{eq:Equbo1} and \eqref{eq:Equbo2} and consider that the original QUBO has no linear terms) and fixing the core spins implies an arbitrary selection of one of the degenerate ground states of the subinstances. Since there are at least two equivalent ground states per community, the probability of fixing all cores according to the global solution is low.
Indeed it is not even guaranteed that the global solution corresponds to a minimum of the original instance when restricted to a single community. 
We remark that the approximation ratio is computed with respect to the original minimum energy, indicating that QAOA's absolute performance is enhanced by the approximated reduction. Finally, as suggested at the end of the previous Section, even better results may be obtained by optimizing over the core spins while keeping all boundary spins fixed to the best assignment found by QAOA for the reduced instance.
 


\section{Discussion}
\label{sec:conclusion}

We have demonstrated the application of divide-and-conquer techniques to solve arbitrary Quadratic Unconstrained Binary Optimization (QUBO) instances by solving smaller Polynomial Unconstrained Binary Optimization (PUBO) instances with fewer variables. To compute the advantage of our proposal, we consider the Max-Cut problem on random 3-regular and 4-regular graphs. While a fully-classical solution based on the exact solver for Max-SAT problem called \texttt{akmaxsat} seems not to benefit from the variable elimination, quantum heuristic algorithms do. In particular, we applied the Quantum Approximate Optimization Algorithm (QAOA) to both the original and reduced instances and find that the latter ones not only require $\sim 42\%$ fewer qubits for random 3-regular graphs (respectively $\sim 22\%$ fewer qubits for 4-regular ones), but also return a considerably improved approximation ratio.

We believe that reaching the scientific and technological goal of demonstrating quantum advantage for practical applications needs to consider quantum and classical processors not as competitors, but as complementary. This view is intrinsic in the formulation of variational quantum algorithms which require several iterations of a classical optimizer, but can be pushed even further. Our proposal uses classical pre-process to manipulate the problem instance and make it more suitable to the quantum algorithm. While classical solvers may also benefit from this pre-process, and this would be a very desirable outcome towards the end goal of solving problems of practical interest, we observe a situation in which the reduced instances were not easier to solve with a certain classical method than the original ones. If confirmed, the different impact of the pre-processing step may contribute to reduce the current performance gap between quantum and classical solvers (despite one cannot rule out the opposite situation). Finally, we expect that post-process may unlock further benefits for quantum algorithms.

In the context of this study, several questions are still open. Just to name a few: Is there a classical solver for the PUBO problem which takes less time on the reduced instances than on the original ones? The time cost strongly depends on the implementation details of the algorithms, can a better implementation change the fractional cost reported in Fig.~\ref{fig:dnc-cost}(right)? QAOA may be used to solve the partially-quenched instances too, what would the performance be in this case? Quantum algorithms for binary optimization is a rich topic and we expect that the stream of exciting results will continue in the next several years.


\section{Methods}
\label{sec:methods}

\subsection{Derivation of the PUBO energy function of a single community}
\label{sec:derivation_pubo}

Here we discuss a constructive method to derive $\tilde{E}^{(c)}_\text{intra-comm}(\vec b_c)$ as defined in Eq.~\eqref{eq:intra-comm-energy}. The method we chose is based on the framework introduced in reference \cite{Sawaya2020}, which has the more ambitious goal of converting Hamiltonians (i.e. quantum energy functions) of arbitrary quantum $d$-level systems to Hamiltonians of quantum spins (or qubits). We will rephrase the method in the context of classical energy functions and spins.


Consider the assignment $\vec b_c$ for the spins in $B_c$. While keeping the boundary spins fixed, we vary the core spins of community $c$ and determine the value of $\tilde{E}^{(c)}_\text{intra-comm}(\vec b_c)$ by minimization. Such value is the energy of assignment $\vec b_c$, but can also be seen as the real number $e_{\vec b_c}$:
\begin{align}
	e_{\vec b_c} \equiv \tilde{E}^{(c)}_\text{intra-comm}(\vec b_c) = \min_{\vec t_c \in \{1,-1\}^{|T_c|}} E^{(c)}_\text{intra-comm}(\vec b_c, \vec t_c) \; .
\end{align}
We repeat a similar minimization for all $2^{|B_c|}$ assignments of the boundary spins and determine all corresponding $e_{\vec b_c}$ values. In an explicit way, it is clear that:
\begin{align}
\label{eq:explicit-e-intra}
	\tilde{E}^{(c)}_\text{intra-comm}(\vec b_c)
&= \sum_{\vec b'_c \in \{1,-1\}^{|B_c|}} e_{\vec b'_c} \, \delta_{\vec b_c, \vec b'_c} \\
&= \sum_{\vec b'_c \in \{1,-1\}^{|B_c|}} e_{\vec b'_c} \, \prod_{i=1}^{|B_c|} \frac{1 + b'_{c, i} b_{c,i}}{2} \; ,
\end{align}
where the Kronecker delta notation has been generalized to vectors and then expressed as product of quadratic factors. By carrying out the products and summation, the standard PUBO form of $\tilde{E}^{(c)}_\text{intra-comm}$ as a polynomial in $b_{c,i}$ is derived.

It is important to comment on the computational cost of this derivation. The above expression has $d=2^{|B_c|}$ terms, each corresponding to $|B_c|$ products of 2-term factors (\textit{i.e.} factors like $\tfrac{1}{2} + \tfrac{b'_{c,i} b_{c,i}}{2}$). If expanded by exhaustive enumeration, we have to sum $d^2$ contributions. In general this cost is much less than that of solving the original problem exhaustively since typically $d^2 \ll 2^N$.
In addition, most of the coefficients typically cancel out, returning a PUBO with a relatively low degree.
%
In the next Section we present a connection to the Walsh-Hadamard transform that allows to reduce the cost from $\mathcal{O}(d^2)$ to $\mathcal{O}(d \log d)$. This corresponds to a logarithmic overhead with respect to the task of finding all the values $e_{\vec b_c}$ since we have $d$ of them.

\subsection{Encoding diagonal operators using Walsh-Hadamard transform}
\label{app:sec:encoding-with-hadamard}

An encoding procedure is required to express a diagonal Hermitian operator as a linear combination
of product of qubit operators. We follow the approach described in \cite{Sawaya2020}, but provide
a more efficient way to compute the coefficients of linear combination.
In the context of this paper, refer to Eq.~\eqref{eq:explicit-e-intra} of the main text and notice that
we are looking for a quantum Hamiltonian of the form:
\begin{equation*}
	\hat H^{(c)}_\text{intra-comm} = \sum_{\vec b \in \{1,-1\}^{|B_c|}} e_{\vec b} \, \ketbra{\vec b}{\vec b} \; .
\end{equation*}
Its eigenstates are trivially given by the computational basis states of $|B_c|$ qubits, and the eigenvalues
are explicitly provided as the $2^{|B_c|}$ real numbers $e_{\vec b}$.
In this Section we use the standard notation adopted by the quantum information community and
rephrase the desired Hamiltonian as an arbitrary, diagonal operator on $M$ qubits:
\begin{equation}
	\hat H = \sum_{\vec s \in \{0,1\}^M} \nu(\vec s) \, \ketbra{\vec s}{\vec s}
\end{equation}

The projector $\ketbra{\vec s}{\vec s}$ can be rewritten as: 
\begin{align}
  \ketbra{\vec s}{\vec s}
&= \prod_{k=0}^{M-1} \left[ \frac{1+(-1)^{s_k} \hat Z_k}{2} \right] \\ 
&= \frac{1}{2^M} \sum_{\vec t \in \{0,1\}^M} \sqrt{2^M} \bra{t} \hat Z_0^{s_0} \hat Z_1^{s_1} \dots \hat Z_{M-1}^{s_{M-1}} \ket{+}^M \,
                                             \hat Z_0^{t_0} \hat Z_1^{t_1} \dots \hat Z_{M-1}^{t_{M-1}} \\
&= \frac{1}{\sqrt{2^M}} \sum_{\vec t \in \{0,1\}^M} \bra{\vec t} \hat Z^{\vec s} \ket{+}^M \, \hat Z^{\vec t} \\
&= \frac{1}{\sqrt{2^M}} \sum_{\vec t \in \{0,1\}^M} (-1)^{\vec t \cdot \vec s} \hat Z_{\vec t} \; ,
\end{align}
where the first line has been obtained using:
\begin{equation}
\label{eq:proj-to-paulis}
\begin{cases}
  \ketbra{0}{0} = (1+\hat Z)/2 \\
  \ketbra{1}{1} = (1-\hat Z)/2 \; ,
\end{cases}
\end{equation}
the third line by adopting the shorthand notation $\hat Z^{\vec s }= \hat Z_0^{s_0} \hat Z_1^{s_1} \dots \hat Z_{N-1}^{s_{N-1}}$
(and similarly for $\hat Z^{\vec t}$), and the fourth by explicit computation and the standard definition of scalar product
$\vec t \cdot \vec s = \sum_{k=0,1,\dots,M-1} t_k \, s_k$.

Substituting inside the expression of $\hat H$, one ontains an expression for the coefficients $f(\vec t)$ of the expansion of $\hat H$ as a linear combination of products of Z Pauli matrices:
\begin{equation}
  H = \sum_{\vec t\in\{0,1\}^M}
      \underbrace{ \sum_{\vec s\in\{0,1\}^M} \frac{\nu(\vec s)}{\sqrt{2^M}} (-1)^{\vec t \cdot \vec s}}_{\equiv f(t)}
      \, \hat Z^{\vec t}
    = \sum_{\vec t\in\{0,1\}^M} f(t) \, \hat Z^{\vec t}
\end{equation}
The definition of $f(\vec t)$ makes it clear that the coefficient of the linear combinations are given by the Walsh-Hadamard transform of the diagonal entries of $\hat H$, namely $\nu(\vec s)$. One can then take advantage of the fast Walsh-Hadamard transform to compute $f(\vec t)$ in time $\mathcal(d \log d)$ with $d$ being the dimensionality of $\nu(\vec s)$, here being $d=2^M$.

\subsection{Reduce PUBO instances to SAT instances}
\label{app:sec:pubo-to-sat}

Polynomial Unconstrained Binary Optimization (PUBO) and Maximum Satisfiability (SAT) instances are important classes of optimization problems involving binary variables. Both classes are NP-complete problems and it is possible to express equivalent instances in either of the two formulations. Here we describe the reduction method we used to convert minimization of PUBO into maximization of SAT. This can be seen as a generalization of the reduction of Max-Cut to Max-2-SAT described in reference \cite{Gramm2003}.

The reduction requires a SAT variable for each PUBO spin and each $k$-body PUBO interaction is translated into $2^{k-1}$ SAT clauses of $k$ variables.

The transformation from PUBO to SAT can be obtained for each term separately. Algorithm\ref{alg:transformation} describes what SAT clauses need to be added to take into account the $k-$body term $s_1 s_2 \dots s_k$. It also updates the quantity \texttt{offset} corresponding to a constant value added to the SAT to faithfully reproduce the PUBO values.

\begin{algorithm}[h]
	\caption{Transform PUBO terms to SAT clauses}
	\label{alg:transformation}
	\begin{algorithmic}[1]
	    \item [The transformation is performed separately for each term of the PUBO instance.]
	    \item [PUBO spins are denoted $\{s_i\}_{i=1,\dots,N}$ while SAT variables are denoted $\{x_i\}_{i=1,\dots,N}$.]
	    \item [Without loss of generality we consider the $k$-body term $s_1 s_2 \dots s_k$.]
	    \item [$C$ denote a SAT clause, $w$ its weight.]
		\State $k \gets$ order of PUBO term
		\State offset $\gets $ same as after transformation of previous term
		\For{$\vec s \in \{+1,-1\}^k$}
			\State $a \gets 1$
			\State $C \gets $ \texttt{True}
			\For{$j = 1, 2, \dots, k$}
				\If{$s_j = +1$}
					\State $C \gets C \lor \lnot x_j$
				\Else
					\State $C \gets C \lor x_j$
					\State $a \gets -a$
				\EndIf
			\EndFor
			\If{$a = -1$}
				\State \textbf{add clause} $C$ with weight $w=2$
			\EndIf
		\EndFor
		\State offset $\gets $ offset $- (2^{k}-1)$ 
	\end{algorithmic}
\end{algorithm}

The algorithm starts with a few initialization: $k$ is the order of the spin term, \texttt{offset} is either 0 or the value after the previous translation. The main loop explores all the $2^k$ assignments of the $k$ spins sequentially (line 3). For each, a tentative clause is constructed as the negative of the spin assignment (line 5:13). The utility variable $a$ (initialized in line 4) computes whether the spin term returns a positive or negative contribution to the PUBO value (line 11). If the contribution is negative, the mirror clause is added to the SAT instance with weight $w=2$ (lines 14:16). After line 17, the clauses added to the SAT have a cumulative value of $w 2^{k-1}=2^k$ for all variable assignments apart from those for which the spin term is negative. In that case, the cumulative value is $2^k-2$.  The offset $(-2^k+1)$ moves the two SAT values to the desired $+1$, respectively $-1$, value.

Apart from a constant offset, the case for $k=2,3$ is presented below for direct inspection:
\begin{align*}
    \text{spins of PUBO: }\{s_1,s_2,\dots ,s_N\} \quad&\rightarrow
        \quad \text{variables of SAT: }\{x_1,x_2, \dots ,x_N\} \, ;\\
    \text{2-body term: }s_1\,s_2 \quad&\rightarrow
        \quad \text{clauses: }(x_1 \lor \lnot x_2)\, , \,(\lnot x_1 \lor x_2) \, ;\\
    \text{3-body term: }s_1\,s_2\,s_3 \quad&\rightarrow
        \quad \text{clauses: }(x_1 \lor x_2 \lor x_3)\, , \,(\lnot x_1 \lor \lnot x_2 \lor x_3)\, , \,
                              (\lnot x_1 \lor x_2 \lor \lnot x_3) \, , \,(x_1 \lor \lnot x_2 \lor \lnot x_3) \, .
\end{align*}

A few observations.
1) If the PUBO term has a non-unit coefficient $c$, all weights for the SAT clauses and the offset are multiplied by $c$.
2) Several SAT solvers search for the maximum value of the SAT instance while PUBO solvers search for the minimum value of the PUBO instance. Inverting the sign of all weights and the offset accounts for this difference.
3) Certain SAT solvers have constraints on the weight values, for example \texttt{akmaxsat} requires the weight to be a positive integer. These constraints can be accounted by a suitable rescaling of the weights and offset. If the coefficient of the PUBO term is negative $c<0$, the condition in line 14 is changed to ``if $a =+1$ then'' and a $(-1)$ factor is added to $w$ and the offset contribution.



\begin{acknowledgments}
The author thanks Jesmin Jahan Tithi for discussion on community detection algorithms and their implementation.
\end{acknowledgments}


\bibliographystyle{unsrt}
\bibliography{references}

\begin{thebibliography}{10}

\bibitem{Kochenberger2014}
Gary Kochenberger, Jin~Kao Hao, Fred Glover, Mark Lewis, Zhipeng L{\"{u}},
  Haibo Wang, and Yang Wang.
\newblock {The unconstrained binary quadratic programming problem: A survey}.
\newblock {\em Journal of Combinatorial Optimization}, 28(1):58--81, 2014.

\bibitem{Glover2018a}
Fred Glover, Gary Kochenberger, and Yu~Du.
\newblock {Quantum Bridge Analytics I: a tutorial on formulating and using QUBO
  models}.
\newblock {\em arXiv:1811.11538}, 2018.

\bibitem{Baxter1982}
Rodney~J. Baxter.
\newblock {\em {Exactly Solved Model in Statistica}}.
\newblock Academic Press, 1982.

\bibitem{Feige1995}
Uriel Feige and Michel Goemanst.
\newblock {Approximating the Value of Two Prover Proof Systems, With
  Applications to AMX 2SAT and MAX DICUT}.
\newblock {\em Proceedings Third Israel Symposium on the Theory of Computing
  and Systems}, pages 182--189, 1995.

\bibitem{Lucas2014}
Andrew Lucas.
\newblock {Ising formulations of many NP problems}.
\newblock {\em Frontiers in Physics}, 2(February):1--15, 2014.

\bibitem{Dunning2018}
Iain Dunning, Swati Gupta, and John Silberholz.
\newblock {What works best when? A systematic evaluation of heuristics for
  Max-Cut and QUBO}.
\newblock {\em INFORMS Journal on Computing}, 30(3):608--624, 2018.

\bibitem{Shor1999}
Peter~W. Shor.
\newblock {Polynomial-time algorithms for prime factorization and discrete
  logarithms on a quantum computer}.
\newblock {\em SIAM Review}, 41(2):303--332, 1999.

\bibitem{Grover1997}
Lov~K. Grover.
\newblock {Quantum mechanics helps in searching for a needle in a haystack}.
\newblock {\em Physical Review Letters}, 79(2):325--328, 1997.

\bibitem{Abrams1999}
Daniel~S. Abrams and Seth Lloyd.
\newblock {Quantum algorithm providing exponential speed increase for finding
  eigenvalues and eigenvectors}.
\newblock {\em Physical Review Letters}, 83(24):5162--5165, dec 1999.

\bibitem{Cao2019}
Yudong Cao, Jonathan Romero, Jonathan~P. Olson, Matthias Degroote, Peter~D.
  Johnson, M{\'{a}}ria Kieferov{\'{a}}, Ian~D. Kivlichan, Tim Menke, Borja
  Peropadre, Nicolas~P.D. Sawaya, Sukin Sim, Libor Veis, and Al{\'{a}}n
  Aspuru-Guzik.
\newblock {Quantum Chemistry in the Age of Quantum Computing}.
\newblock {\em Chemical Reviews}, 119(19):10856--10915, 2019.

\bibitem{Kadowaki1998}
Tadashi Kadowaki and Hidetoshi Nishimori.
\newblock {Quantum annealing in the transverse Ising model}.
\newblock {\em Physical Review E}, 58(5):5355, 1998.

\bibitem{Farhi2001a}
Edward Farhi, Jeffrey Goldstone, Sam Gutmann, Joshua Lapan, Andrew Lundgren,
  and Daniel Preda.
\newblock {A quantum adiabatic evolution algorithm applied to random instances
  of an NP-complete problem}.
\newblock {\em Science}, 292(5516):472--475, apr 2001.

\bibitem{Boixo2014}
Sergio Boixo, Troels~F. R{\o}nnow, Sergei~V. Isakov, Zhihui Wang, David Wecker,
  Daniel~A. Lidar, John~M. Martinis, and Matthias Troyer.
\newblock {Evidence for quantum annealing with more than one hundred qubits}.
\newblock {\em Nature Physics}, 10(February):218, 2014.

\bibitem{Santra2014}
Siddhartha Santra, Gregory Quiroz, Greg {Ver Steeg}, and Daniel~A. Lidar.
\newblock {Max 2-SAT with up to 108 qubits}.
\newblock {\em New Journal of Physics}, 16(4):045006, apr 2014.

\bibitem{Venturelli2015}
Davide Venturelli, Salvatore Mandr{\`{a}}, Sergey Knysh, Bryan O'Gorman, Rupak
  Biswas, and Vadim~N. Smelyanskiy.
\newblock {Quantum optimization of fully connected spin glasses}.
\newblock {\em Physical Review X}, 5:031040, 2015.

\bibitem{Farhi14_qaoa_orig}
Edward Farhi, Jeffrey Goldstone, and Sam Gutmann.
\newblock {A quantum approximate optimization algorithm}.
\newblock {\em arXiv:1411.4028}, 2014.

\bibitem{Wecker16_training}
D.~{Wecker}, M.~B. {Hastings}, and M.~{Troyer}.
\newblock {Training a quantum optimizer}.
\newblock {\em Phys. Rev. A}, 94(2):022309, 2016.

\bibitem{Guerreschi17_qaoa_opt}
G.~{Giacomo Guerreschi} and M.~{Smelyanskiy}.
\newblock {Practical optimization for hybrid quantum-classical algorithms}.
\newblock arXiv:1701.01450, 2017.

\bibitem{Zhou2018}
Leo Zhou, Sheng-Tao Wang, Soonwon Choi, Hannes Pichler, and Mikhail~D. Lukin.
\newblock {Quantum Approximate Optimization Algorithm: Performance, Mechanism,
  and Implementation on Near-Term Devices}.
\newblock {\em Physical Review X}, 10(2):021067, 2018.

\bibitem{Shaydulin2019a}
Ruslan Shaydulin, Ilya Safro, and Jeffrey Larson.
\newblock {Multistart Methods for Quantum Approximate Optimization}.
\newblock {\em 2019 IEEE High Performance Extreme Computing Conference, HPEC
  2019}, 2019.

\bibitem{Streif2019a}
Michael Streif and Martin Leib.
\newblock {Training the Quantum Approximate Optimization Algorithm without
  access to a Quantum Processing Unit}.
\newblock {\em arXiv:1908.08862}, 2019.

\bibitem{Otterbach2017a}
J.~S. Otterbach, R.~Manenti, N.~Alidoust, A.~Bestwick, M.~Block, B.~Bloom,
  S.~Caldwell, N.~Didier, E.~{Schuyler Fried}, S.~Hong, P.~Karalekas, C.~B.
  Osborn, A.~Papageorge, E.~C. Peterson, G.~Prawiroatmodjo, Nick Rubin, C.~A.
  Ryan, D.~Scarabelli, M.~Scheer, E.~A. Sete, P.~Sivarajah, R.~S. Smith,
  A.~Staley, N.~Tezak, W.~J. Zeng, A.~Hudson, B.~R. Johnson, M.~Reagor, M.~P.
  da~Silva, and Chad~T. Rigetti.
\newblock {Unsupervised machine learning on a hybrid quantum computer}.
\newblock {\em arXiv:1712.05771}, 2017.

\bibitem{Mueller2018}
Peter M{\"{u}}ller, Marc Ganzhorn, Andreas Fuhrer, Stefan Filipp, Kristan
  Temme, Andrew Cross, Ivano Tavernelli, Abhinav Kandala, Jay~M. Gambetta, John
  Smolin, Lev~S. Bishop, Walter Riess, Gian Salis, Panagiotis Barkoutsos,
  Antonio Mezzacapo, Daniel~J. Egger, Nikolaj Moll, and Jerry~M. Chow.
\newblock {Quantum optimization using variational algorithms on near-term
  quantum devices}.
\newblock {\em Quantum Science and Technology}, 3:030503, 2018.

\bibitem{Pichler2018a}
Hannes Pichler, Sheng-tao Wang, Leo Zhou, Soonwon Choi, and Mikhail~D. Lukin.
\newblock {Quantum optimization for maximum independent set using Rydberg atom
  arrays}.
\newblock {\em arXiv:1808.10816}, 2018.

\bibitem{Pagano2020}
Guido Pagano, Aniruddha Bapat, Patrick Becker, Katherine~S. Collins, Arinjoy
  De, Paul~W. Hess, Harvey~B. Kaplan, Antonis Kyprianidis, Wen~Lin Tan,
  Christopher Baldwin, Lucas~T. Brady, Abhinav Deshpande, Fangli Liu, Stephen
  Jordan, Alexey~V. Gorshkov, and Christopher Monroe.
\newblock {Quantum approximate optimization of the long-range Ising model with
  a trapped-ion quantum simulator}.
\newblock {\em Proceedings of the National Academy of Sciences of the United
  States of America}, 117(41):25396--25401, 2020.

\bibitem{google2020b}
Frank Arute, Kunal Arya, Ryan Babbush, Dave Bacon, Joseph~C. Bardin, Rami
  Barends, Sergio Boixo, Michael Broughton, Bob~B. Buckley, David~A. Buell,
  Brian Burkett, Nicholas Bushnell, Yu~Chen, Zijun Chen, Ben Chiaro, Roberto
  Collins, William Courtney, Sean Demura, Andrew Dunsworth, Edward Farhi,
  Austin Fowler, Brooks Foxen, Craig Gidney, Marissa Giustina, Rob Graff, Steve
  Habegger, Matthew~P. Harrigan, Alan Ho, Sabrina Hong, Trent Huang, L.~B.
  Ioffe, Sergei~V. Isakov, Evan Jeffrey, Zhang Jiang, Cody Jones, Dvir Kafri,
  Kostyantyn Kechedzhi, Julian Kelly, Seon Kim, Paul~V. Klimov, Alexander~N.
  Korotkov, Fedor Kostritsa, David Landhuis, Pavel Laptev, Mike Lindmark,
  Martin Leib, Erik Lucero, Orion Martin, John~M. Martinis, Jarrod~R. McClean,
  Matt McEwen, Anthony Megrant, Xiao Mi, Masoud Mohseni, Wojciech Mruczkiewicz,
  Josh Mutus, Ofer Naaman, Matthew Neeley, Charles Neill, Florian Neukart,
  Hartmut Neven, Murphy~Yuezhen Niu, Thomas~E. O'Brien, Bryan O'Gorman, Eric
  Ostby, Andre Petukhov, Harald Putterman, Chris Quintana, Pedram Roushan,
  Nicholas~C. Rubin, Daniel Sank, Kevin~J. Satzinger, Andrea Skolik, Vadim
  Smelyanskiy, Doug Strain, Michael Streif, Kevin~J. Sung, Marco Szalay, Amit
  Vainsencher, Theodore White, Z.~Jamie Yao, Ping Yeh, Adam Zalcman, and Leo
  Zhou.
\newblock {Quantum Approximate Optimization of Non-Planar Graph Problems on a
  Planar Superconducting Processor}.
\newblock {\em arXiv:2004.04197}, 2020.

\bibitem{Guerreschi2019}
Gian~Giacomo Guerreschi and Anne~Y. Matsuura.
\newblock {QAOA for Max-Cut requires hundreds of qubits for quantum speed-up}.
\newblock {\em Scientific Reports}, 9:6903, 2019.

\bibitem{Dalzell2020}
Alexander~M. Dalzell, Aram~W. Harrow, Dax~Enshan Koh, and Rolando~L. la~Placa.
\newblock {How many qubits are needed for quantum computational supremacy?}
\newblock {\em Quantum}, 4:264, 2020.

\bibitem{Bravyi2020}
Sergey Bravyi, Alexander Kliesch, Robert Koenig, and Eugene Tang.
\newblock {Obstacles to Variational Quantum Optimization from Symmetry
  Protection}.
\newblock {\em Physical Review Letters}, 125:260505, 2020.

\bibitem{Kugel2012}
Adrian K{\"{u}}gel.
\newblock {Improved exact solver for the weighted Max-SAT problem}.
\newblock {\em Proc. Pragmatics of SAT Workshop (POS-10)}, 8:15--27, 2012.

\bibitem{Karp1972}
Karp~R. M.
\newblock Reducibility among combinatorial problems.
\newblock In R.~E. Miller, J.~W. Thatcher, and J.~D. Bohlinger, editors, {\em
  Complexity of Computer Computations. The IBM Research Symposia Series}, pages
  85--103. Springer, Boston, MA, 1972.

\bibitem{Barkoutsos2020}
Panagiotis~Kl. Barkoutsos, Giacomo Nannicini, Anton Robert, Ivano Tavernelli,
  and Stefan Woerner.
\newblock {Improving Variational Quantum Optimization using CVaR}.
\newblock {\em Quantum}, 4:256, 2020.

\bibitem{Larkin2020a}
Jason Larkin, Mat{\'{i}}as Jonsson, Daniel Justice, and Gian~Giacomo
  Guerreschi.
\newblock {Evaluation of Quantum Approximate Optimization Algorithm based on
  the approximation ratio of single samples}.
\newblock {\em arXiv:2006.04831}, 2020.

\bibitem{Fortunato2016}
Santo Fortunato and Darko Hric.
\newblock {Community detection in networks: A user guide}.
\newblock {\em Physics Reports}, 659:1--44, 2016.

\bibitem{igraph}
Gabor Csardi and Tamas Nepusz.
\newblock The igraph software package for complex network research.
\newblock {\em InterJournal}, Complex Systems:1695, 2006.

\bibitem{Sawaya2020}
Nicolas P.~D. Sawaya, Tim Menke, Thi~Ha Kyaw, Sonika Johri, Al{\'{a}}n
  Aspuru-Guzik, and Gian~Giacomo Guerreschi.
\newblock {Resource-efficient digital quantum simulation of d-level systems for
  photonic, vibrational, and spin-s Hamiltonians}.
\newblock {\em npj Quantum Information}, 6:49, 2020.

\bibitem{Goemans1995}
M.~X. Goemans and D.~P. Williamson.
\newblock {Improved approximation algorithms for maximum cut and satisfiability
  problems using semidefinite programming}.
\newblock {\em Journal Of The Association for Computing Machines},
  42(6):1115--1145, 1995.

\bibitem{Halperin2004}
Eran Halperin, Dror Livnat, and Uri Zwick.
\newblock {MAX CUT in cubic graphs}.
\newblock {\em Journal of Algorithms}, 53(2):169--185, 2004.

\bibitem{Larson2018}
Jeffrey Larson and Stefan~M. Wild.
\newblock {Asynchronously parallel optimization solver for finding multiple
  minima}.
\newblock {\em Mathematical Programming Computation}, 10(3):303--332, 2018.

\bibitem{2020SciPy-NMeth}
Pauli Virtanen, Ralf Gommers, Travis~E. Oliphant, Matt Haberland, Tyler Reddy,
  David Cournapeau, Evgeni Burovski, Pearu Peterson, Warren Weckesser, Jonathan
  Bright, St{\'e}fan~J. {van der Walt}, Matthew Brett, Joshua Wilson, K.~Jarrod
  Millman, Nikolay Mayorov, Andrew R.~J. Nelson, Eric Jones, Robert Kern, Eric
  Larson, C~J Carey, {\.I}lhan Polat, Yu~Feng, Eric~W. Moore, Jake
  {VanderPlas}, Denis Laxalde, Josef Perktold, Robert Cimrman, Ian Henriksen,
  E.~A. Quintero, Charles~R. Harris, Anne~M. Archibald, Ant{\^o}nio~H. Ribeiro,
  Fabian Pedregosa, Paul {van Mulbregt}, and {SciPy 1.0 Contributors}.
\newblock {{SciPy} 1.0: Fundamental Algorithms for Scientific Computing in
  Python}.
\newblock {\em Nature Methods}, 17:261--272, 2020.

\bibitem{Gramm2003}
Jens Gramm, Edward~A. Hirsch, Rolf Niedermeier, and Peter Rossmanith.
\newblock {Worst-case upper bounds for MAX-2-SAT with an application to
  MAX-CUT}.
\newblock {\em Discrete Applied Mathematics}, 130:139--155, 2003.

\end{thebibliography}

\end{document}